\documentstyle[pre,aps,multicol,epsfig]{revtex}

\begin{document}

\begin{abstract}
\end{abstract}

\begin{center}
{\large Domain Walls in Two-Component Dynamical Lattices} \vskip0.5truecm

P.G. Kevrekidis$^{1}$, Boris A. Malomed$^{2}$, D.J. Frantzeskakis$^{3}$, and
A.R. Bishop$^{4}$
\end{center}

$^{1}${\small Department of Mathematics and Statistics, University of
Massachusetts, Amherst MA 01003-4515, USA}

$^{2}${\small Department of Interdisciplinary Studies, Faculty of
Engineering, Tel Aviv University, Tel Aviv, Israel}

$^{3}${\small Department of Physics, University of Athens,
Panepistimiopolis, Zografos, Athens 15784, Greece}

$^{4}${\small Center for Nonlinear Studies and Theoretical Division, Los
Alamos National Laboratory, Los Alamos, NM 87545 USA}

\bigskip \bigskip

\begin{center}
Abstract
\end{center}

We introduce domain-wall (DW) states in the bimodal discrete nonlinear 
Schr{\"{o}}dinger equation, in which the modes are coupled by cross phase
modulation (XPM). The results apply to an array of nonlinear optical
waveguides carrying two different polarizations of light, or two different
wavelengths, with anomalous intrinsic diffraction controlled by direction of
the light beam, and to a string of drops of a binary Bose-Einstein
condensate, trapped in an optical lattice. By means of continuation from
various initial patterns taken in the anti-continuum (AC) limit, we find a
number of different solutions of the DW type, for which different stability
scenarios are identified. In the case of strong XPM coupling, DW
configurations contain a single mode at each end of the chain. The most
fundamental solution of this type is found to be always stable. Another
solution, which is generated by a different AC pattern, demonstrates
behavior which is unusual for nonlinear dynamical lattices: it is unstable
for small values of the coupling constant $C$ (which measures the ratio of
the nonlinearity and coupling lengths), and becomes stable at larger $C$.
Stable bound states of DWs are also found. DW configurations generated by
more sophisticated AC patterns are identified as well, but they are either
completely unstable, or are stable only at small values of $C$. In the case
of weak XPM, a natural DW solution is the one which contains a combination
of both polarizations, with the phase difference between them $0$ and $\pi $
at the opposite ends of the lattice. This solution is unstable at all values
of $C$, but the instability is very weak for large $C$, indicating
stabilization as the continuum limit is approached. The stability of DWs is
also verified by direct simulations, and the evolution of unstable DWs is
simulated too; in particular, it is found that, in the weak-XPM system, the
instability may give rise to a moving DW. The DW states can be observed
experimentally in the same parameter range where discrete solitons have been
found in the lattice setting.

\newpage

\section{Introduction}

Nonlinear optical fibers and waveguides and arrays composed of them furnish
a unique example of a medium in which various solitary-wave patterns and
their complexes can be easily observed in a real experiment, and described,
with a very high accuracy, by relatively simple models, a benchmark 
example being the
nonlinear Schr\"{o}dinger equation \cite{b24}. Besides commonly known bright
and dark solitons, solitary-wave structures in the form of domain walls
(DWs) were also predicted in a fiber with normal group-velocity dispersion
(GVD) which carries two different waves with orthogonal polarizations,
circular or linear, that interact through the cross-phase modulation (XPM)
induced by the Kerr nonlinearity \cite{Wab,Malomed}. Similar structures can
also be expected to exist in planar nonlinear optical waveguides 
\cite{Haelt}. DW patterns are distinguished by the 
property that, asymptotically, they
contain a single polarization, with a switch between two of them in a
localized region. In fact, solutions for the optical DWs were constructed,
following the pattern of earlier known solutions of the DW type for a system
of coupled Ginzburg-Landau equations that describe interactions between roll
patterns with different orientations in a convection layer \cite{MNT,Alik}.

GVD in the fiber must be normal in order to prevent the modulational
instability (MI) of the DW's uniform background fields. Nevertheless, it is
known that a uniform two-component field, unlike single-component ones, may
be subject to MI even in the case of normal GVD \cite{b24}. Loosely, DWs are
related to the MI of the two-component uniform state the same way as the
usual bright soliton is related to MI of the single-component field in the
case of anomalous GVD \cite{Haelt}.

The same model (which is presented in detail below) applies to an altogether
different physical system, namely, a string of drops of a binary
(two-component) Bose-Einstein condensate (BEC) trapped at minima of a
periodic potential, which can be readily induced by an optical interference
pattern \cite{opticalLattice}. In this connection, it is relevant to mention
that stable DW configurations have been predicted in a continuous
quasi-one-dimensional (cigar-shaped) binary BEC \cite{BECdw}.

Optical DWs in nonlinear fibers have been observed in direct experiments 
\cite{DWexperiment}, including high-repetition periodic DW trains \cite
{DWarray}. On the other hand, recent experimental achievements in the
observation of discrete spatial optical solitons in arrays of waveguides in
the spatial domain \cite{Yaron,anomalous} suggest that observation of
DW-like structures in waveguide arrays may be quite feasible too.
Additionally, solitons of the DW type may be a new species of solitary waves
in the discrete nonlinear Schr\"{o}dinger (DNLS) equations, which have
recently attracted a great deal of interest (for a recent review see e.g.,
Ref. \cite{review}). Given that the solutions to coupled DNLS equations have
been examined for a considerable while now (the first relevant results 
appeared
about 20 years ago in \cite{eilb}), it appears that 
DWs may be one of the few types of DNLS
solitons that have not been studied yet. 

An objective of the present work is to introduce this type of discrete
solitons and study their stability by means of precise numerical methods. An
exact formulation of the model, together with estimates of relevant physical
parameters, are given in section 2. In section 3, we study in detail DWs in
the case of strong XPM, which corresponds to circular polarizations. We find
several different types of DWs, the simplest one being stable for all values
of the inter-site coupling constant $C$ (which is the ratio of the
propagation length determined by the Kerr nonlinearity to the
linear-coupling length, in terms of the optical-waveguide array). DWs of a
different type exhibit a rather unusual stability behavior, being stable at
small values of $C$, and stable at larger $C$, i.e., in weakly and strongly
coupled arrays, respectively. The existence of stable bound states of DWs is
also demonstrated. Other types of DWs turn out to be either completely
unstable, or stable only at small values of $C$. In section 4, we consider
DWs in the model with weak XPM, which corresponds to linear polarizations.
In this case, DWs are unstable. However, the instability growth rate of the
simplest (fundamental) DW becomes vanishingly small for large values of $C$,
so that the pattern becomes (marginally) stable in the continuum limit, 
$C\rightarrow \infty $. In all the cases, the predicted stability of DWs is
tested in direct numerical simulations, and in those cases when DWs are
expected to be unstable, the instability development is simulated too.
Section 5 summarizes the paper.

\section{Formulation of the model}

The model of an array of nonlinear optical fibers carrying fields $\phi
_{n}(z)$ and $\psi _{n}(z)$, which correspond to two orthogonal
polarizations of light, has the form 
\begin{eqnarray}
i\frac{d}{dz}\left( \psi _{n}\right)  &=&C\left( \psi _{n+1}+\psi
_{n-1}-2\psi _{n}\right) -(|\psi _{n}|^{2}+\beta |\phi _{n}|^{2})\psi _{n}\,,
\label{dweq1} \\
i\frac{d}{dz}\left( \phi _{n}\right)  &=&C\left( \phi _{n+1}+\phi
_{n-1}-2\phi _{n}\right) -(|\phi _{n}|^{2}+\beta |\psi _{n}|^{2})\phi _{n}\,,
\label{dweq2}
\end{eqnarray}
where $z$ is the propagation distance along the fiber and $n$ is the index
of the lattice site. Equations (\ref{dweq1}) and (\ref{dweq2}) are written
in a rescaled form, in which the constant $C$ of the linear coupling between
adjacent fibers has a straightforward physical meaning: it is a ratio of the
characteristic propagation length $L_{{\rm nonlin}}$ along the waveguide,
determined by the Kerr nonlinearity (self-phase modulation, SPM), to the
coupling length $L_{{\rm coupl}}$ determined by the linear interaction
between adjacent waveguides. In the most typical experimental situation,
with the power of light beams $500-1000$ W, the nonlinearity length in {\rm
AlGaAs} waveguides (whose nonlinearity is $500$ time stronger than that of
fused silica glass) takes values $\sim 1-2$ mm, and the coupling length is 
$\sim 1$ mm \cite{Yaron,anomalous}. In fact, $L_{{\rm coupl}}$ may vary in
broad limits, as it exponentially depends on the separation $h$ between the
waveguides: for instance, $L_{{\rm coupl}}$ decreases by a factor of $1.6$
as $h$ increases from $9$ $\mu $m to $11$ $\mu $m \cite{Yaron}. Results
displayed below (such as a width of DWs in terms of the number of lattice
sites) suggest that stable DWs in optical-waveguide arrays should be
observable in the same experimental setups, and in essentially the same
range of the power-per-waveguide, where bright discrete solitons have been
found in works \cite{Yaron,anomalous}.

In\ Eqs. (\ref{dweq1}) and (\ref{dweq2}), the SPM coefficient is normalized
to be $1$, and $\beta $ is a relative coefficient of the XPM nonlinear
coupling, which is $2$ or $2/3$, in the case of linear or circular
polarizations, respectively. In the latter case, we neglect, as usual,
four-wave-mixing (FWM) nonlinear coupling terms \cite{b24} Strictly
speaking, for the short propagation distance ($\sim 10$ mm), relevant to the
recent experiments \cite{Yaron}, it may be necessary to keep the FWM terms
in the case $\beta =2/3$; we do not consider this issue in the present work
because, as it will be shown below, most interesting results are found in
the case $\beta =2$. We will refer to the two cases with $\beta =2/3$ and 
$\beta =2$ as those with the weak and strong XPM coupling, respectively.
The value $\beta =2$ applies also to the case when the two modes refer not
to polarizations, but rather to light signals carried by different
wavelengths \cite{b24}, which is another plausible realization of the
present model in terms of nonlinear optics.

If the nonlinearity in the waveguides is induced by the usual self-focusing
Kerr effect, and the light is launched so that its Poynting vector is
oriented parallel to the array, the linear coupling between fibers ({\it
discrete diffraction}) corresponds to $C<0$ in Eqs. (\ref{dweq1}) and (\ref
{dweq2}). In this case, the system gives rise to two-component discrete
bright solitons, which were recently studied in detail, including their
generalization to the two-dimensional lattice \cite{us}. However, a newly
developed experimental technique, based on launching an oblique beam into
the array \cite{anomalous}, makes it possible to implement {\em anomalous}
discrete diffraction, which corresponds to $C>0$ in Eqs. (\ref{dweq1}) and 
(\ref{dweq2}). In this case, bright solitons do not exist.

As it was mentioned above, the same system of Eqs. (\ref{dweq1}) and (\ref
{dweq2}) may also be regarded as a normalized model of a string of
binary-BEC drops trapped in an optical lattice. In\ this case, assuming the
usual situation with the positive scattering length (i.e., repulsion between
atoms), one has $C>0$, which is precisely what is necessary to generate DW
solutions. Unlike the realization in terms of nonlinear optics, the value of
the coefficient $\beta $ may be arbitrary (but positive). The separation
between the drops is of the order of the lattice-generating light
wavelength, i.e., $\sim 1$ $\mu $m. A promising candidate for the binary
condensate being a mixture of $^{85}$Rb and $^{87}$Rb, the corresponding
positive scattering lengths being $\sim 10$ nm \cite{Rb}. The necessary
temperature and densities of the drops can be estimated as $10^{-5}-10^{-6}$
K and $10^{11}-10^{12}$ cm$^{-3}$, which can be readily achieved by
available experimental techniques

In this work, we examine discrete domain walls (DWs) in the system (\ref
{dweq1}) and (\ref{dweq2}), which may be stable only if $C>0$. We will
construct DW solutions, starting from the anti-continuum (AC) limit with 
$C=0 $ \cite{MacAub}, and using the continuation in $C$ to extend the
solutions up to bifurcation points where they lose 
their stability, or up to
the continuum limit ($C\rightarrow \infty $) when possible. Thus, we seek
solutions of the form 
\begin{eqnarray}
\psi _{n} &=&\exp (i\Lambda z)u_{n}\,,  \label{dweq3} \\
\phi _{n} &=&\exp (i\Lambda z)v_{n},  \label{dweq4}
\end{eqnarray}
arriving at the stationary equations 
\begin{eqnarray}
F(u_{n},v_{n})\equiv -C\Delta _{2}u_{n}+(|u_{n}|^{2}+\beta
|v_{n}|^{2})u_{n}-\Lambda u_{n} &=&0,  \label{dweq5} \\
G(u_{n},v_{n})\equiv -C\Delta _{2}v_{n}+(|v_{n}|^{2}+\beta
|u_{n}|^{2})v_{n}-\Lambda v_{n} &=&0.  \label{dweq6}
\end{eqnarray}
Once a solution to Eqs. (\ref{dweq5}) and (\ref{dweq6}) has been found (by
means of a Newton-type numerical method), we perform the linear stability
analysis around it, looking for perturbed solutions as \cite{Eil3} 
\begin{eqnarray}
\psi _{n} &=&\exp (i\Lambda z)\left[ u_{n}+\epsilon a_{n}\exp (i\omega
z)+\epsilon b_{n}\exp (-i\omega ^{\star }z)\right] ,  \label{dweq7} \\
\phi _{n} &=&\exp (i\Lambda z)\left[ v_{n}+\epsilon c_{n}\exp (i\omega
z)+d_{n}\exp (-i\omega ^{\star }z)\right] \,  \label{dweq8}
\end{eqnarray}
(the asterisk stands for the complex conjugation), and solving the ensuing
matrix eigenvalue problem.

\section{Numerical results for the strong-XPM model}

In this section, we consider the model based on Eqs. (\ref{dweq1}) and (\ref
{dweq2}) with $\beta =2$. We start the examination of DW structures by
looking for stationary solutions in a natural form which is taken, in the AC
limit, as 
\begin{equation}
u_{n}=(...0,0,0,V_{1},1,1,1,...),\,v_{n}=(...1,1,1,V_{2},0,0,0,...)
\label{DW}
\end{equation}
[see Eqs. (\ref{dweq3}) and (\ref{dweq4})], where $V_{1}$ and $V_{2}$ belong
to one (central) site of the lattice and will be defined below. We will, in
particular, consider the steady states with $\Lambda =1$ in Eqs. 
(\ref{dweq3}) and (\ref{dweq4}), but results will be 
generally true if $1$ in AC {\it
ansatz} (\ref{DW}) is replaced by $\sqrt{\Lambda }$ for an arbitrary
positive value of $\Lambda $. Notice that in the present study, we fix
$\Lambda$ and vary $C$, modifying, essentially, in this way the degree
of localization or equivalently the peak power. One can 
instead always rescale $C$ to the value $C=1$, and equivalently
vary the propagation constant $\Lambda$.

\begin{figure}[tbp]
\epsfxsize=8.5cm \centerline{\epsffile{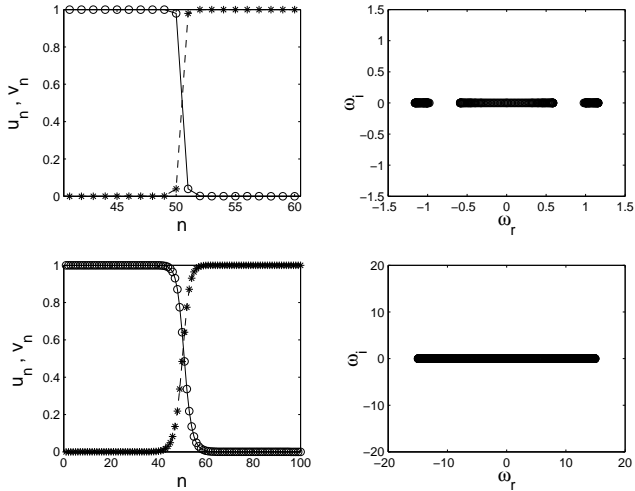}}
\caption{Left panels show examples of the domain-wall solutions generated,
in the case $\protect\beta=2$ and $\Lambda=1$, by the AC pattern with 
$u_n=(0,\dots,0,1,1,\dots,1)$ (the dashed line connecting the stars is as a
guide to the eye) and $v_n=(1,\dots,1,0,0, \dots,0)$ (circles connected by
the solid line). Right panels show the spectral plane ($\protect\omega_r,
\protect\omega_i$) of the corresponding stability eigenfrequencies found
from the equations linearized about the stationary solution, the subscripts
standing for the real and imaginary parts. The absence of eigenfrequencies
with a nonzero imaginary part indicates the stability of the configuration.
The top and bottom panels correspond, respectively, to $C=0.034$ (weak
coupling) and $C=3.5$ (strong coupling).}
\label{dwf1}
\end{figure}

Numerical calculations demonstrate that, for $\beta =2$ (which corresponds
to the circular polarizations of light in the fibers), the AC pattern (\ref
{DW}) with 
\begin{equation}
V_{1}=1,\,V_{2}=0  \label{robust}
\end{equation}
(or vice versa) generates the most structurally robust and stable solutions.
In particular, this solution was found to be stable for {\em all} values of
the coupling constant $C$, up to the continuum limit $C\rightarrow \infty $.
Examples of such a DW for cases of weak ($C=0.034$) and strong ($C=3.5$)
inter-site coupling (recall $C$ may be realized as the ratio $L_{{\rm nonlin}
}/L_{{\rm coupl}}$ in the optical-waveguide array) are shown in Fig. \ref
{dwf1}. The complete stability of DWs belonging to this branch of the
solutions has also been verified by direct simulations of the full nonlinear
equations (\ref{dweq1}) and (\ref{dweq2}), with initial conditions
containing a small noisy component.

The next case to consider is when 
\begin{equation}
V_{1}=V_{2}=\sqrt{\Lambda /(1+\beta )}  \label{vector}
\end{equation}
in Eq. (\ref{DW}). These values imply that, in the AC limit, the central
waveguide in the array carries a vectorial state, in which both polarities
have equal amplitudes. The numerical investigation reveals an unusual
feature of this solution branch: it is unstable for small values of $C$,
starting from the AC limit ($C=0$), but becomes {\em stable} at $C=0.61$,
and remains stable thereafter up to $C=\infty $. This property is opposite
to the common scenario, when a sufficiently strong discreteness is
responsible (through the effective potential energy barrier that it creates)
for the stabilization of various solitary-wave lattice patterns (for
instance, two-dimensional pulses \cite{kim} or vortices \cite{vortex}) that
are unstable in the continuum limit.

\begin{figure}[tbp]
\epsfxsize=8.5cm \centerline{\epsffile{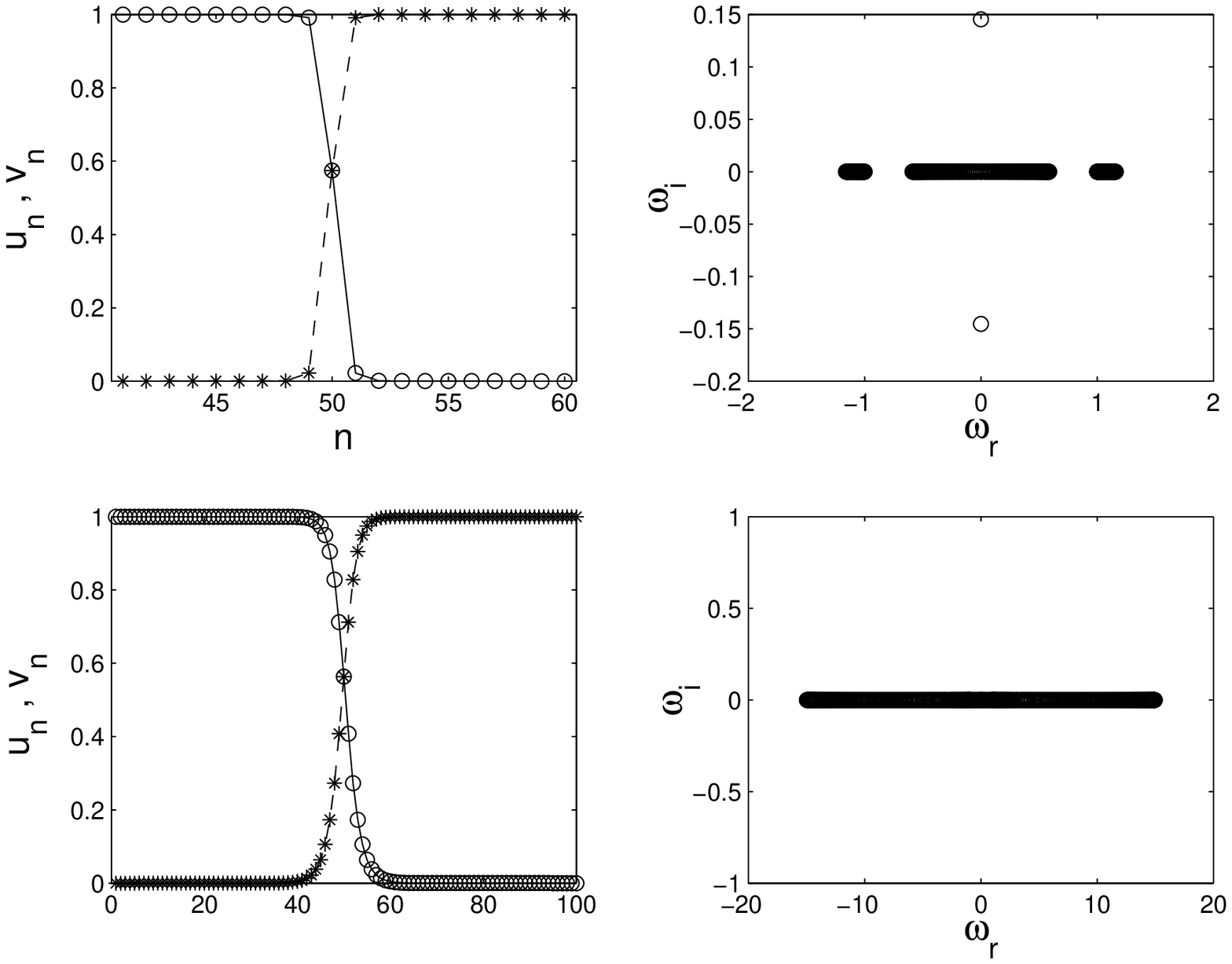}} \epsfxsize=8.5cm 
\centerline{\epsffile{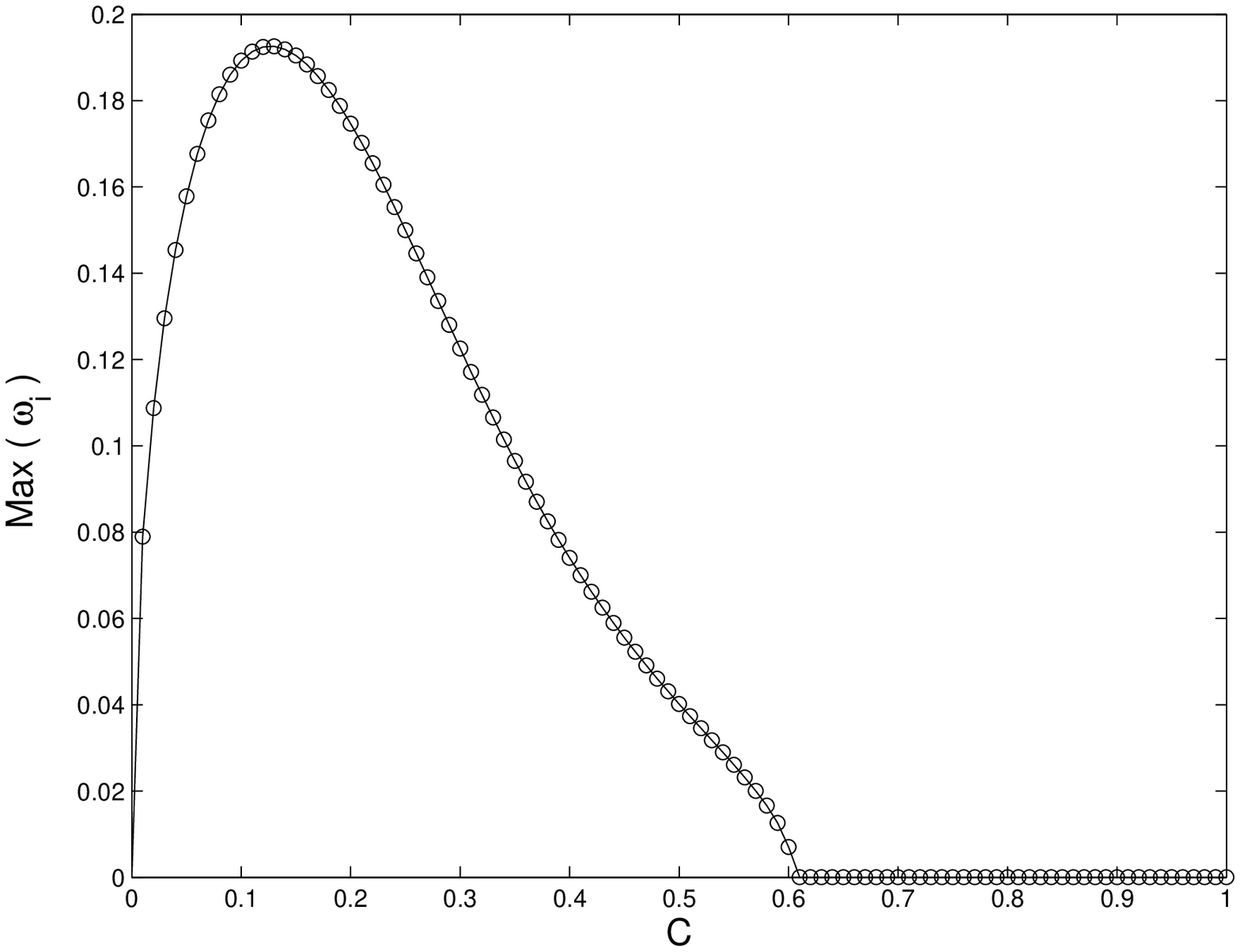}}
\caption{The two top rows show DW solutions and their stability eigenvalues,
as generated by the AC pattern (\ref{DW}) at the same values of parameters
as in Fig. 1, but with $V_1$ and $V_2$ chosen as in Eq. (\ref{vector}). The
bottom panel shows the imaginary part of the most unstable eigenvalue as a
function of $C$.}
\label{dwf2}
\end{figure}

Solutions generated by Eq. (\ref{vector}) are shown, for the same values of 
$\beta $, $\Lambda $, and $C$ as in Fig. \ref{dwf1}, in two upper rows of
Fig. \ref{dwf2} (top panel). Notice the presence of an unstable (imaginary)
eigenvalue pair in the panel pertaining to $C=0.034$. The imaginary part of
the unstable eigenvalue is displayed, as a function of the coupling
constant, in the lower panel of Fig. \ref{dwf2}, showing the transition from
instability to stability at $C\approx 0.61$.

\begin{figure}[tbp]
\epsfxsize=8.5cm \centerline{\epsffile{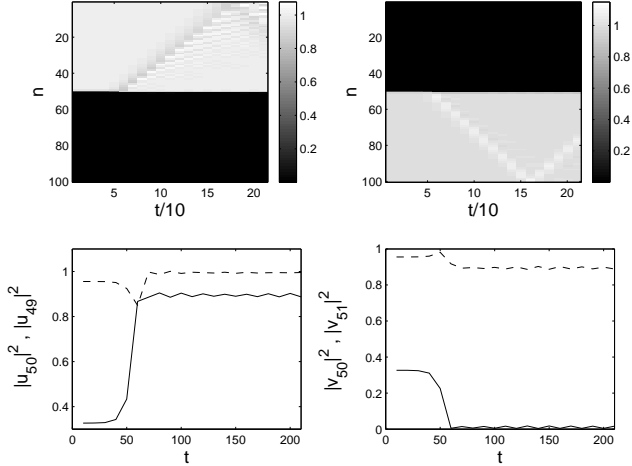}}
\caption{Evolution of an unstable DW configuration belonging to the solution
branch generated by the AC ansatz (\ref{DW}) and (\ref{vector}) 
with $C=0.1$, $\protect\beta = 2$, and $\Lambda =1$. 
To initiate the onset of the
instability, a noise with an amplitude $10^{-4}$ was added to the initial
configuration. The left and right top panels display the evolution of the
fields $|v_n|$ and $|u_n|$, while the bottom panels show the evolution of
the fields at the sites $n=50$ (solid curve) and $n=49$ (dashed curve) for 
$u $, and $n=50$ (solid curve) and $n=51$ (dashed) for $v$.}
\label{dw2}
\end{figure}

The predictions for the stability of these solutions were also checked
against direct simulations of the full equations. In the case when the DW is
predicted to be stable, it is indeed
found to be completely stable (not
shown here). In the case when it is expected to be unstable, the simulations
show (see Fig. \ref{dw2}) that the unstable DW emits a packet of lattice
phonons and rearranges itself so that the field in one component at one site
($n=50$) in Fig. \ref{dw2}) becomes equal to the field in the other
component at the adjacent site ($n=51$, in Fig. \ref{dw2}). Comparison with
the expressions (\ref{DW}) and (\ref{robust}) clearly suggests that the
result of the instability development is the rearrangement of the DW into a
stable one belonging to the solution branch generated by Eq. (\ref{robust})
in the AC limit.

\begin{figure}[tbp]
\epsfxsize=8.5cm \centerline{\epsffile{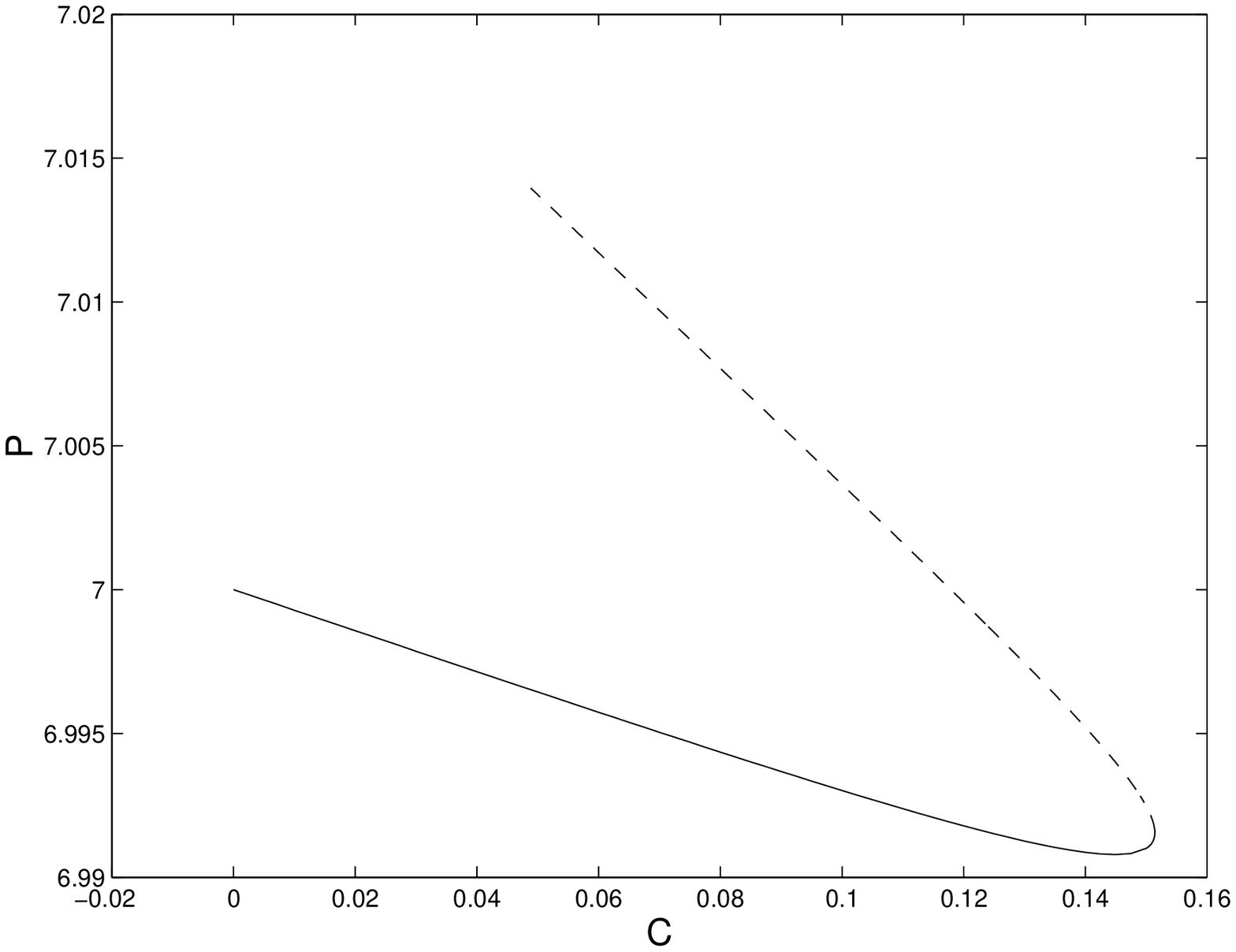}} \epsfxsize=8.5cm 
\centerline{\epsffile{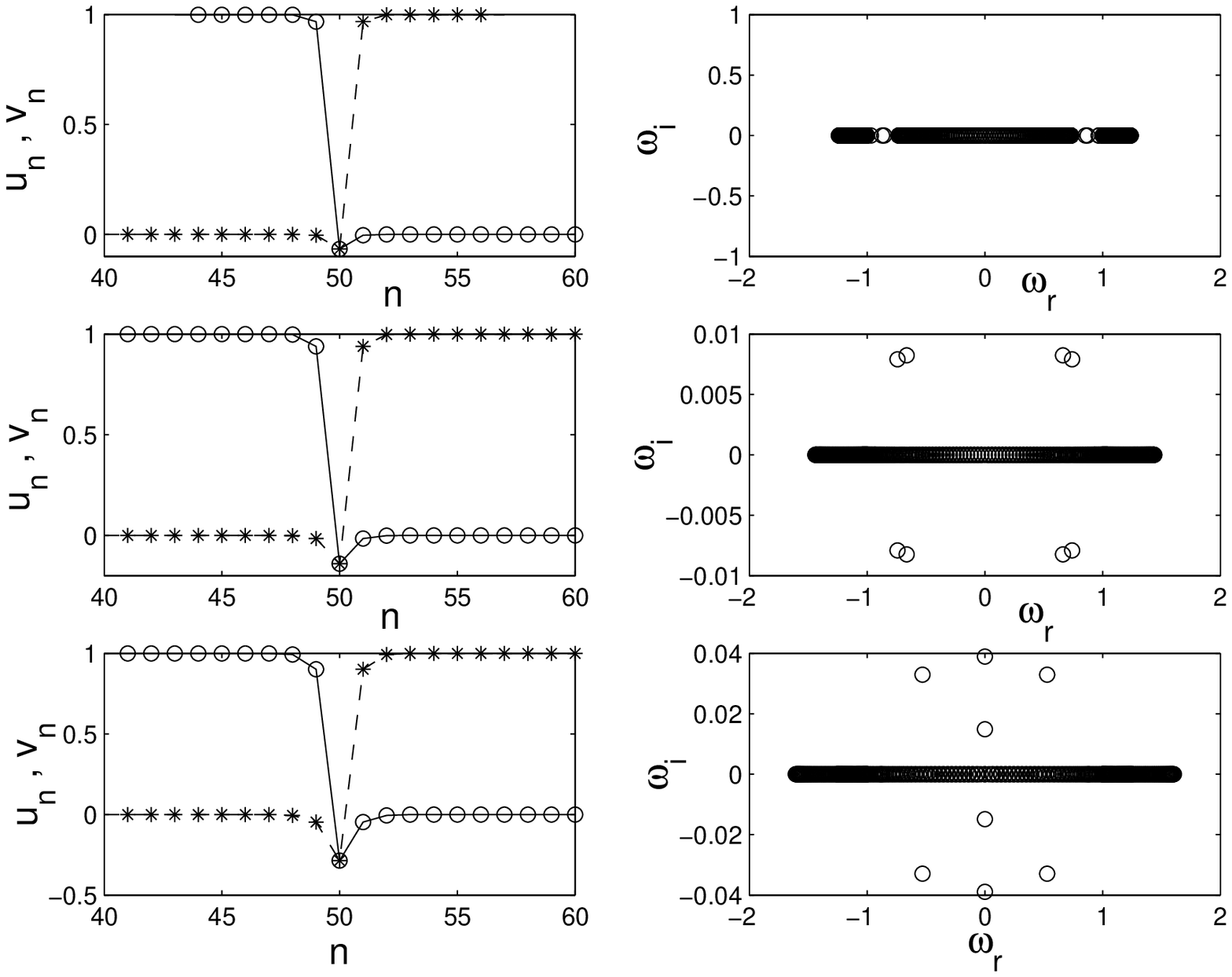}}
\caption{The upper panel demonstrates the result of continuation of the
branch with $V_1=V_2=0$ from the AC limit up to the turning point ($C\approx
0.15143$) -- the solid curve -- and back -- the dashed one. To this end, the 
$L^2$ norm $P$ (see Eq. (\ref{P})) of one component (the other behaves
similarly) of the solution is shown as a function of the coupling constant 
$C $. The lower panel shows solutions at three points along the solid line in
the upper panel. In particular, the stable solution for $C=0.06$ (its
spatial profile and stability) is shown in the top subplot, and solutions
for $C \approx 0.11$ and $C \approx 0.15142$ (the latter one being very
close to the turning point) are displayed in the middle and bottom subplots.}
\label{dwf3}
\end{figure}

A solution generated by the AC pattern (\ref{DW}) with $V_{1}=V_{2}=0$
(i.e., with both field components equal to zero at the central lattice site
in the AC limit) was also examined. Unlike the solution branches considered
above, this one does not reach the continuum limit. Instead, it terminates
at $C=0.15143$ through a turning-point bifurcation. This is obvious in Fig. 
\ref{dwf3}, which displays the $L^{2}$ norm of one component of the solution
(the other component behaves similarly), 
\begin{equation}
P\equiv \left( \sum_{-\infty }^{+\infty }u_{n}^{2}\right) ^{1/2}\,,
\label{P}
\end{equation}
vs. the coupling constant $C$ (note that $P$ is finite due to the finiteness
of the computational domain). The dependence $P(C)$, shown in 
Fig. \ref{dwf3}, is not a completely invariant 
characteristic, as values of $P$ depend on
the size of the domain. However, the turning-point structure is invariant.

In the lower panel of Fig. \ref{dwf3}, the present solution and its linear
stability eigenvalues are shown for three different values of $C$ ($C=0.06$, 
$C=0.11$, and $C=0.15143$ in the in the top, middle, and bottom subplots).
As is seen from this part of the figure, the first solution is stable, while
the latter two are not. The solutions of the present type are stable for 
$C<0.0725$, and the spectrum of their linear-stability eigenvalues contains
two continuous bands. As $C$ increases, two pairs of eigenvalues bifurcate
from the outer band and move towards the inner one. Their first collision
with the inner band occurs at $C\approx 0.0725$, and the second collision
takes place at $C\approx 0.08$. Each collision generates an
instability-bearing quartet of eigenvalues, manifesting the so-called
Hamiltonian Hopf bifurcation \cite{vdm}. Subsequently (for larger $C$), the
two quartets move towards the imaginary eigenvalue axis. Very close to the
turning point, the collision of the first quartet with the axis induces a
symmetry breaking, which results in one pair of eigenvalues moving towards
the origin, while another pair moves upwards along the imaginary axis.
Finally, as a result of the collapse of the lower imaginary pair onto the
origin of the spectral plane, the turning-point bifurcation occurs and the
branch terminates, as is shown in the upper panel of Fig. \ref{dwf3}.

\begin{figure}[tbp]
\epsfxsize=8.5cm \centerline{\epsffile{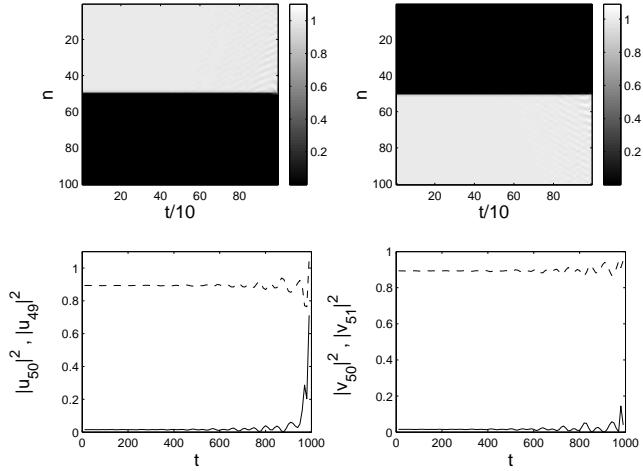}}
\caption{ The evolution of a weakly unstable solution generated, in the AC
limit, by the ansatz (\ref{DW}) with $V_1=V_2=0$, for $C=0.1$, $\protect
\beta = 2$, $\Lambda = 1$. The development of the instability is initiated
by adding noise with an initial amplitude $10^{-3}$. The meaning of the
panels is the same as in Fig. \ref{dw2}.}
\label{dw5}
\end{figure}

Direct simulations again show that the solutions which are stable according
to the linearization are indeed stable  in full simulations. An example of
direct simulations of the unstable solution belonging to this branch is
given in Fig. (\ref{dw5}). As is seen, the 
instability does set in, but
its growth is extremely slow.

We also considered the situation with the zero-field state occupying, in the
AC limit, {\em two} (or more) lattice sites. The resulting scenario turns
out to be similar to that demonstrated above for the patterns generated by
the AC limit (\ref{DW}) with $V_{1}=V_{2}=0$: the solution is stable at very
small $C$, then becomes unstable due to the bifurcation of four (instead of
two in the previous case) pairs of eigenvalues from the outer band and their
collision with the inner band (not shown here). Finally, the branch
terminates at a turning point at $C\approx 0.17$, similarly to what was
shown in Fig. \ref{dwf3}. Note that this turning point is found at a larger
value of $C$ than in Fig. \ref{dwf3}. We continued the analysis, starting
with several empty (zero-field) sites in the center of the pattern at $C=0$,
which produced quite a similar picture, an unexpected feature being that the
value of $C$ at the turning point {increases} with the increase of the
number of the initially empty sites.

\begin{figure}[tbp]
\epsfxsize=8.5cm \centerline{\epsffile{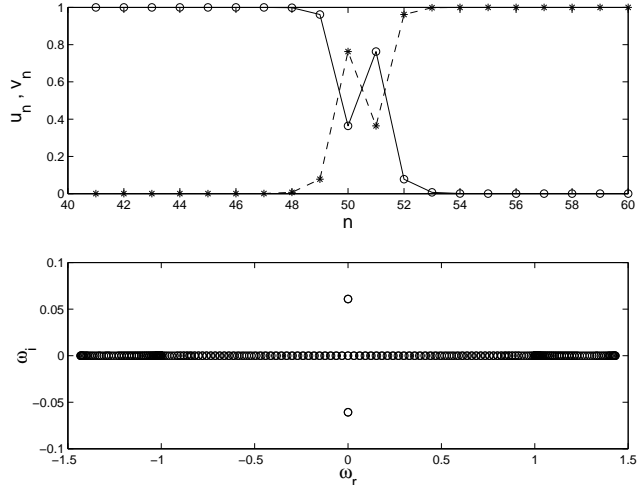}}
\caption{The solution generated, in the AC limit, by the pattern with two
sites occupied by the vectorial state (\ref{vector}) is shown exactly at the
turning point of $C \approx 0.1085$, along with its stability eigenvalues.
This solution branch is always unstable, as discussed in the text.}
\label{dwf4}
\end{figure}

Contrary to what was described above for the case of one or several empty
sites, in the case when the initial pattern has two sites occupied by the
vectorial state (\ref{vector}), the solution behaves quite differently from
its counterpart in the case when only one site was initially occupied by the
vectorial state. In fact, not only does the branch terminate - in this case,
at $C\approx 0.1085$ (the solution found at the turning point and its
stability eigenvalues are shown in Fig. \ref{dwf4}) - but it is also found
to be {\em always unstable}: for every value of $C$ starting from the AC
limit $C=0$, there are two imaginary eigenvalue pairs. The branch terminates
when one of them passes through the origin.

\begin{figure}[tbp]
\epsfxsize=8.5cm \centerline{\epsffile{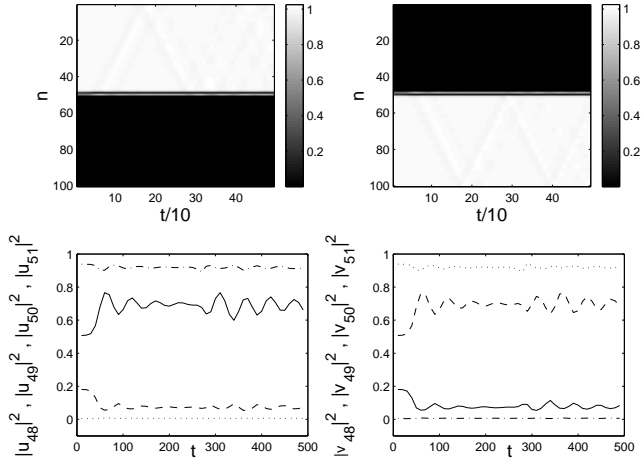}}
\caption{The evolution of an unstable solution generated, in the AC limit,
by the ansatz (\ref{DW}) with two sites occupied by the vectorial state (\ref
{vector}), for $C=0.1$, $\protect\beta = 2$, $\Lambda = 1$. The development
of the instability is initiated by adding noise with an initial amplitude 
$10^{-4}$. The meaning of the panels is the same as in Fig. \ref{dw2}, the
bottom ones showing (absolute value of) 
the field evolution at the points $n=48$ (dash-dotted
lines), $n=49$ (dashed lines), $n=50$ (solid lines), and $n=51$ (dotted
lines) for $u$ (bottom left) and $v$ (bottom right) as a function of
time $t$.}
\label{dw6}
\end{figure}

An example of directly simulated evolution of an unstable DW of this type is
shown in Fig. \ref{dw6}. In this case, the instability development is
essentially faster than in the case shown in Fig. \ref{dw2}, and, as a
result, a stable DW appears.

\begin{figure}[tbp]
\epsfxsize=8.5cm \centerline{\epsffile{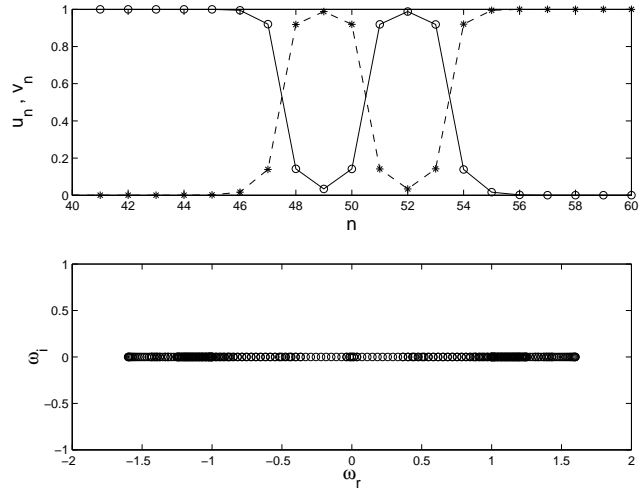}}
\caption{An example of a stable complex consisting of three domain walls for 
$C=0.15$. Shown are the solution's spatial profile (top panel) and stability
eigenvalues (in the bottom panel). }
\label{dwf5}
\end{figure}

Furthermore, stable DW solutions found above, such as the ones generated by
the AC-limit pattern (\ref{DW}) with $V_{1}$ and $V_{2}$ chosen as per Eqs. 
(\ref{robust}) or (\ref{vector}), can form {\em stable} complexes
(higher-order DWs). Obviously, the complex must contain an odd number of
fundamental DWs. An example is shown in Fig. \ref{dwf5} for $C=0.15$. Direct
simulations confirm that these complexes are completely stable.

Recall that, if the underlying system of Eqs. (\ref{dweq1}) and (\ref{dweq2})
is applied to the string of BEC drops, the value of $\beta$, which is
determined by the values of the cross-scattering length, is arbitrary. In
the case $\beta >1$, the results are quite similar to those described in
detail above for $\beta =2$. For $\beta < 1$, the results are very
different, as is described in the next section.

\section{The weak-XPM model}

The underlying pattern (\ref{DW}), used to construct all the DW states
considered above, was suggested by solutions found (in the temporal, rather
than spatial, domain) in the continuum model of the single nonlinear optical
fiber \cite{Malomed,Haelt} or cigar-shaped BEC \cite{BECdw}. Solutions of
this type in the continuum model exist only in the case of sufficiently
strong XPM, namely, for $\beta >1$.

Unlike the continuum model, in the discrete system (\ref{dweq1}), (\ref
{dweq2}) DW patterns of the type (\ref{DW}) can also be found for $\beta <1$
(weak XPM coupling), including the case $\beta =2/3$, which is specially
relevant for the applications to nonlinear optics. However, in this case
properties of the DW solutions are drastically different from those
described above for $\beta =2$.

\begin{figure}[tbp]
\epsfxsize=8.5cm \centerline{\epsffile{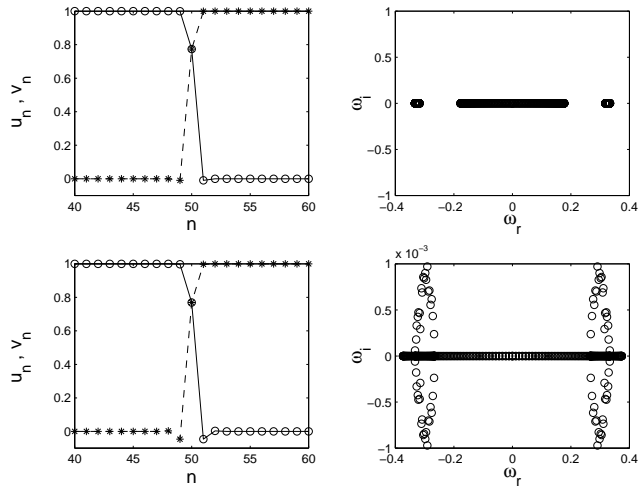}}
\caption{The DW solution and the corresponding stability eigenvalues prior
to the onset of the oscillatory instabilities (for $C=0.004$), and after the
onset (for $C=0.017$), in the case $\protect\beta =2/3$ and $\Lambda =1$.}
\label{dwf6}
\end{figure}

For $\beta =2/3$, all solutions of the DW type were observed to become {\it
staggered} when continued from the AC limit (i.e., the phase difference
between the fields at adjacent sites of the lattice tends to be $\pi $,
rather than $0$). As a result, the outer continuous band of the stability
eigenvalues has the Krein signature \cite{Aubry} opposite to that of the
inner band, and as soon as the bands collide (which happens at $C\approx
0.011$, i.e., still in the case of very weak coupling), numerous quartets of
complex eigenvalues arise. An example, corresponding to the solution
generated by the AC pattern (\ref{DW}) with the central site occupied by the
vectorial state (\ref{vector}) (with $\beta =2/3$) is displayed in Fig. \ref
{dwf6}. Similar results were obtained for the solution branches generated by
the pattern (\ref{DW}) with $V_{1}=1$ 
and $V_{2}=0$ (also with $\beta =2/3$).
Thus, discrete DW solutions of this type may exist in the weak-XPM case,
unlike the continuum limit, but they are stable only at very small values of
the coupling constant. These conclusions, concerning the existence and
(in)stability of the DW patterns for the weak-XPM case, are corroborated by
direct simulations; however, the instability is extremely slow, therefore it
is not shown here.

In the regime of weak XPM, another type of DW solutions is suggested by
analogy to the findings of Ref. \cite{Malomed} for the continuum version of
the model. This type of solution exists only for the weak-XPM case ($\beta
<1 $) in the continuum model, and, quite naturally, turns out to be
primarily relevant in the same case in the discrete model. In the continuum
limit, this solution is 
\begin{equation}
u(x)=\sqrt{\Lambda }\cos (\chi ),\,v(x)=\sqrt{\Lambda }\sin (\chi ),
\label{dweq10}
\end{equation}
where $\Lambda $ is the same as in Eqs. (\ref{dweq3}) and (\ref{dweq4}),
and, in the first approximation (which corresponds to a broad DW), 
\begin{equation}
\chi (x)=\frac{\pi }{4}-\tan ^{-1}\left[ \exp (-s\sqrt{\epsilon \Lambda }x)
\right] ,\,\epsilon \equiv 1-\beta .  \label{chi}
\end{equation}

\begin{figure}[tbp]
\epsfxsize=8.5cm \centerline{\epsffile{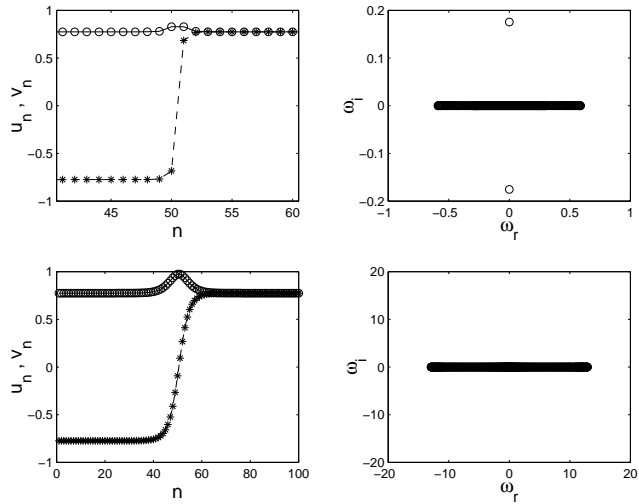}} \epsfxsize=8.5cm
\caption{The upper panel shows the discrete counterpart of the
continuum-model solutions (\ref{dweq10}) for $C = 0.034$ (top
panel) and $C = 3$ (bottom panel). The stability eigenvalues for both
configurations are shown in the respective right subplots. The lower panel
shows the $L^2$ norm of the two branches (the solid and dashed lines
correspond to the norms of the fields $u_n$ and $v_n$, respectively), and
the imaginary part of the most unstable eigenvalue vs. $C$.}
\label{dwf7}
\end{figure}

The corresponding AC limit of such solutions is (for $\Lambda =1$) 
\begin{equation}
u_{n}=(1/\sqrt{2},\dots ,1/\sqrt{2},1/\sqrt{2},\dots ,1/\sqrt{2}
),\,v_{n}=(-1/\sqrt{2},\dots ,-1/\sqrt{2},1/\sqrt{2},\dots ,1/\sqrt{2}).
\label{weakXPM}
\end{equation}
Starting from Eq. (\ref{weakXPM}), the solution was extended all the way
from $C=0$ to the continuum limit, which shows that the solution exists for
all values of $C$. The upper panel of Fig. \ref{dwf7} shows two examples of
this solution, for $C=0.034$ and $C=3$. The lower panel of the figure
clearly shows that this DW state is unstable at all values of $C$, but gets
stabilized in the continuum limit, $C\rightarrow \infty $. For instance, for 
$C=3$ the growth rate of the relevant instability is already extremely
small, $\sim 10^{-6}$, hence this state may seem a practically stable one in
the discrete case too, provided that the coupling constant is large enough.

\begin{figure}[tbp]
\epsfxsize=8.5cm \centerline{\epsffile{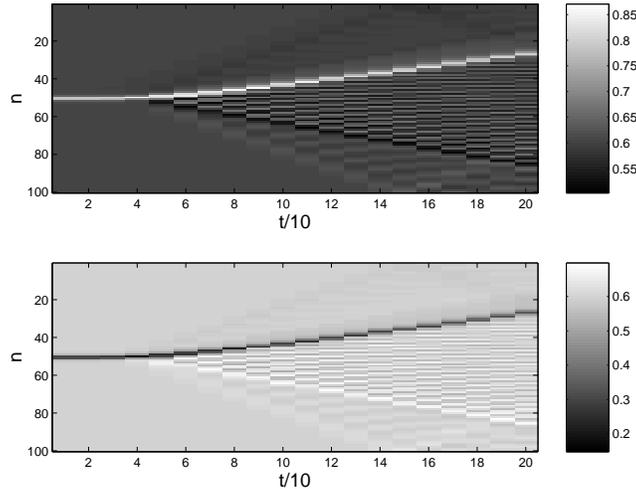}}
\caption{The (norm contour plot of the) 
development of the instability of the DW solution generated by
the ansatz (\ref{dweq10}) in the AC limit in the weak-XPM case, $\protect
\beta = 2/3$, and $C=0.1$, $\Lambda=1$. The instability is initiated by
adding noise with an initial amplitude $10^{-4}$.}
\label{dw10}
\end{figure}

These predictions for the DW patterns generated by the ansatz 
(\ref{weakXPM}) were verified in direct simulations of the 
full nonlinear system with 
$\beta =2/3$. Figure \ref{dw10} shows the development of the relatively
strong instability in the case $C=0.1$. A remarkable result, which makes
this case drastically different from those considered above, is that the
instability transforms the quiescent DW into a {moving} one, which is
accompanied by emission of quasi-linear lattice waves.

\begin{figure}[tbp]
\epsfxsize=8.5cm \centerline{\epsffile{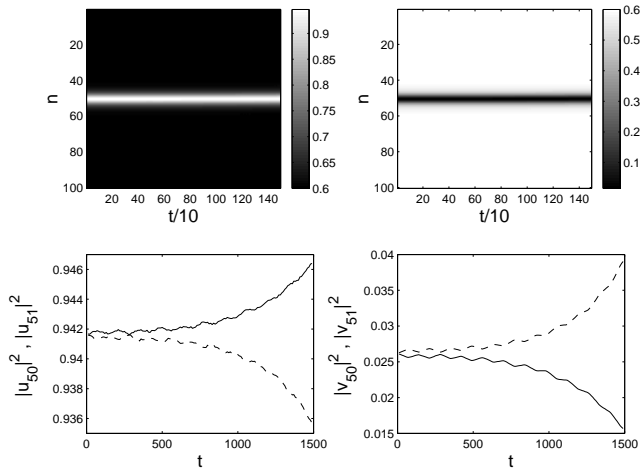}}
\caption{The instability of the DW solution generated by the ansatz (\ref
{dweq10}) in the AC limit in the weak-XPM case, $\protect\beta = 2/3$, and 
$C=1$, $\Lambda=1$. The instability is initiated by adding noise with an
initial amplitude $10^{-4}$. The upper panels show the evolution of the
fields $|u_n|$ and $|v_n|$, while the lower ones specially show the
evolution of the fields at the sites $n=50$ and $n=51$, by means of the
solid and dashed curves, respectively. As it is obvious from the lower
panels, the growth of the instability is extremely slow.}
\label{dw11}
\end{figure}

It has also been verified that, in full accordance with the prediction of
the linear-stability analysis, the formally unstable DWs of the present type
seems to be virtually stable ones. A typical example of that is displayed in
Fig. \ref{dw11}.

\begin{figure}[tbp]
\epsfxsize=8.5cm \centerline{\epsffile{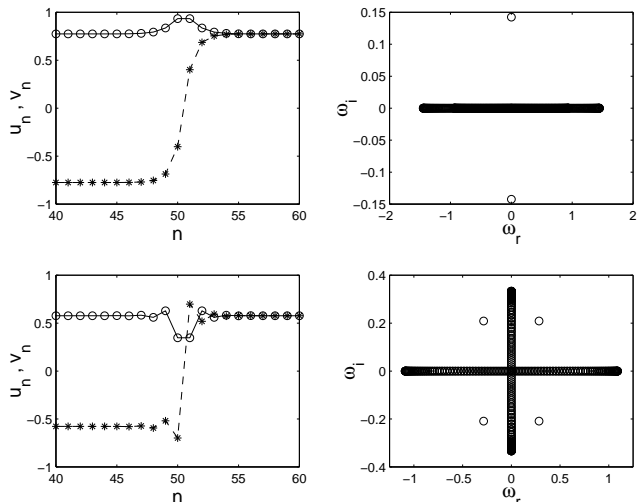}}
\caption{The solution of the type (\ref{dweq10})-(\ref{chi}) is shown for 
$C=0.2$ in the top panel, along with its stability eigenvalues. For
comparison, the bottom panel shows a solution of the same type for $\protect
\beta=2$, $C=0.12$. Notice the continuous-spectrum-induced instabilities in
the latter case.}
\label{dwf8}
\end{figure}

Even though the continuum-limit analytical expressions (\ref{dweq10}) - (\ref
{chi}) for these solutions are only valid for $\beta <1$, one can seek for
similar solutions in the discrete strong-XPM system, with $\beta =2$. The
same way as solutions of the type (\ref{DW}) can exist in the discrete
system with $\beta =2/3$, despite the fact that they do not exist in the
continuum limit for the weak-XPM case (see above), the solutions of the type
(\ref{dweq10}) - (\ref{chi}) have a chance to exist in the discrete
strong-XPM model. We have found that such solutions indeed exist for very
small values of $C$, but they exhibit very strong instabilities, with a part
of the imaginary eigenvalue axis being populated by the continuous spectrum.
In Fig. \ref{dwf8}, a solution of this type for $\beta =2$ and $C=0.12$ is
compared with a ``natural'' one existing at $\beta =2/3$ and $C=0.2$.

\begin{figure}[tbp]
\epsfxsize=8.5cm \centerline{\epsffile{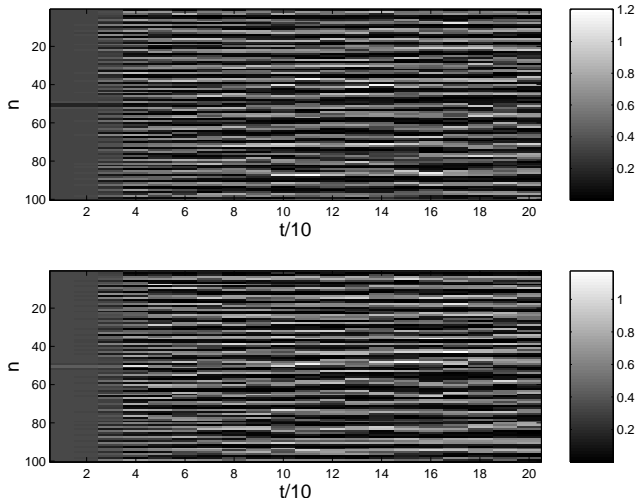}}
\caption{The instability of the DW solution generated by the ansatz (\ref
{dweq10} in the AC limit in the strong-XPM case, $\protect\beta = 2$, and 
$C=0.1$, $\Lambda=1$. The instability is initiated by adding noise with an
initial amplitude $10^{-4}$, leading very rapidly to lattice 
turbulence. The panels show the time evolution of the
(contours of the squares of the) 
fields $|u_n|$ and $|v_n|$.}
\label{dw12}
\end{figure}

The nonlinear evolution of this strong instability was directly simulated,
showing a quick transition to a state of ``lattice turbulence", see Fig. \ref
{dw12}. Thus, the lattice admits the existence of the ``unnatural" solutions
in both strong- and weak-XPM cases, but it never 
allows them to persist ad infinitum.

\section{Conclusions}

In this work we have studied, by means of numerical methods, the structure
and stability of domain-wall (DW) solutions in the system of two discrete
nonlinear Schr{\"{o}}dinger equations with the coupling of the cross phase
modulation (XPM) type. The consideration of this problem is suggested by the
analogy with known DW solutions in a standard model of a nonlinear optical
fiber carrying two polarizations of light or two different wavelengths, as
well as in the quasi-one-dimensional binary Bose-Einstein condensate (BEC).
The results directly apply to an array of fibers of this type, with the
anomalous intrinsic diffraction controlled by the direction of the light
beam, or to a string of BEC drops trapped in an optical-lattice potential;
in the latter case, the generic case with the positive scattering lengths is
that which may give rise to DW patterns.

Using Newton-type methods and continuation from various initial patterns,
starting from the anti-continuum (AC) limit, we have found a number of
different stationary solutions of the DW type (while the continuum model
admits a single type of the DW solution). Different stability scenarios and
transitions to instability were identified for these solutions. In the case
of strong XPM coupling, corresponding to two circular polarizations or two
different wavelengths in the optical-waveguide array, natural DW
configurations contain only one polarization at each end of the chain. The
most fundamental solution of this type, generated by the simplest AC
pattern, has been found to be always stable. Another solution, generated by
the AC pattern that includes a vectorial state in the central site of the
lattice, generates a behavior which is unusual for nonlinear dynamical
lattices: it is unstable for small values of the coupling constant $C$
(which is the ratio of the nonlinear propagation length to the coupling
length in the waveguide array), acquiring stability and remaining stable at
larger values of $C$. Stable bound states formed by stable DWs were also
found. A number of DW configurations generated by more sophisticated AC
patterns were obtained, but they were either found to be completely
unstable, or to be stable only at very small values of $C$.

In the case of weak XPM, which corresponds to linear polarizations in
optics, the solutions become staggered and are subject to oscillatory
instabilities. In this case, a more natural DW solution is the one with a
combination of both polarizations, with the phase difference between them
being $0$ and $\pi $ at the opposite ends of the lattice. This solution is
unstable at all values of $C$, but the instability is very weak for large
values of $C$, corresponding to stabilization in the continuum limit.

The robustness of all the patterns that were predicted to be stable in
direct simulations was corroborated in direct simulations of the full
nonlinear system. The evolution of unstable patterns was simulated too. In
some cases, this instability is extremely weak; in other cases, a stronger
instability leads to rapid rearrangement of the unstable DW into a stable one,
including the possibility of generation of a moving DW in the weak-XPM
model; a very strong instability may even induce ``lattice turbulence".

Estimates of physical parameters necessary for the formation of DWs in the
optical-waveguide array and BEC string were given too. In particular,
discrete optical DWs are expected to be found in the same region of
parameters were bright discrete solitons have been already observed.

The consideration of DW patterns in dynamical lattices can be continued in
several directions. A topic of direct interest concerns the mobility of DWs
across the lattice. Note, in particular, that a mechanical twist applied to an
optical fiber may give rise to a driving force acting on DWs in it \cite
{Malomed}, which can support the motion. On the other hand, a weak
symmetry-breaking deformation of the waveguides will induce linear mixing
between the two orthogonal polarizations, which will drastically affect 
the DWs.
Another interesting problem is the interaction between DWs (in Ref. \cite
{Malomed}, it was shown that a bound state of two DWs with opposite
polarities is possible in a nonlinear optical fiber in the presence of the
twist). These issues will be considered elsewhere.

%\begin{references}

\end{document}